%% file: main.tex
\documentclass[a4paper,11pt]{article}
\usepackage{jheppub} 
\usepackage{lineno}
\usepackage{main-defs}
\usepackage{array}
\usepackage{booktabs}

\title{\boldmath Ultra Fast Calorimeter Simulation\\with Generative Machine Learning on FPGAs}

\author[a]{P. Alex May}
\author[b]{Qibin Liu}
\author[b]{Julia Gonski}
\author[b]{Benjamin Nachman}
\affiliation[a]{San Jose State University, 1 Washington Square, San Jose, CA 95192, USA}
\affiliation[b]{SLAC National Accelerator Laboratory, 2575 Sand Hill Rd, Menlo Park, CA 94025, USA}

\emailAdd{alex.may@sjsu.edu}

\abstract{
Computationally expensive, high-accuracy detector simulations are a major bottleneck for many particle physics experiments such as those at the Large Hadron Collider (LHC) as well as those planned for future colliders.  This challenge has motivated the development of fast generative machine learning based surrogates. We present a hardware-aware variational autoencoder model for fast calorimeter simulation that is designed specifically for field programmable gate array (FPGA) deployment, offering faster and lower power inference capability. Quantization aware training and other compression techniques are applied to respect the resource constraints of a single FPGA. The synthesized implementation of the VAE decoder achieves sub-millisecond latency, resulting in a substantial speed up compared to a traditional GPU implementation with only a small performance drop. This feasibility study demonstrates the potential of utilizing existing FPGA architecture at the LHC and other facilities for efficient offline computing using online resources.
}

\begin{document}
\maketitle
\flushbottom

\input{sections/intro}
\input{sections/methods}
\input{sections/results}

\input{sections/conclusions}

\acknowledgments

PAM is supported by the U.S. Department of Energy under contract number DE-SC0024518.
JG, QL, and BN are supported by the U.S. Department of Energy under contract number DE-AC02-76SF00515.

\appendix

\section{Code Availability}
The machine learning code used in this study is publicly available at \href{https://github.com/SLAC-Julia-Group/CaloGen-VAE-FPGA}{CaloGen-VAE-FPGA}.

\input{sections/app_hls4ml}


\bibliographystyle{JHEP}
\bibliography{biblio.bib}


\end{document}

%% file: sections/intro.tex
\section{Introduction}
\label{sec:intro}


The Monte Carlo (MC) simulation of physics processes is essential for all scientific programs in particle physics. 
High quality and high statistics simulations are necessary for nearly all aspects of data analysis and future planning, including event reconstruction, background estimation, and uncertainty quantification. 
As the number of recorded events increases, experiments require correspondingly larger simulated datasets in order to match the statistical precision of recorded data. 
Full detector simulation based on \textsc{Geant4}~\cite{AGOSTINELLI2003250} provides high fidelity but is computationally expensive.  For example, at the Large Hadron Collider (LHC), a majority of current computing resources are used for MC simulation, and this is projected to increase nearly exponentially in the High Luminosity LHC era~\cite{CERN-LHCC-2022-005,Software:2815292}.
Other HEP experiments such as detection of internally reflected Cherenkov light (DIRC) detectors face a similar situation, where the computational cost grows with the increasing demands of high-granularity simulations and multi-dimensional design optimization~\cite{Hardin_2016,KALICY2024169168}.
In particular, calorimeter shower simulation dominates the cost, accounting for approximately 80\% of total full-simulation time~\cite{Aad_2022} 
This computational burden represents a significant bottleneck for producing the large simulated samples required for future LHC operations.               

Fast simulation techniques can mitigate these computational challenges by reducing the cost per simulated event while maintaining sufficient accuracy for physics analyses.
One approach is the use of parameterized detector response, replacing the propagation of incident particles inside the calorimeter volume by directly generating energy deposits based on a detector parametrization~\cite{Aad_2022,sc2025,Abdullin_2011,sekmen2016cmsfast}. 
Another increasingly viable approach is the use of generative machine learning (ML)~\cite{Paganini:2017hrr,Paganini_2018,Hashemi_2024,Krause_2025,Feickert:2021ajf}. 
Such models learn the relationship between particle inputs and detector outputs from fully simulated data, enabling expensive per-particle propagation to be replaced by a single neural-network inference, thus significantly decreasing simulation latency while retaining essential detector features.
Modern generative ML simulation techniques have become sufficiently performant to be deployed to model the ATLAS~\cite{Aad_2022,ATLAS:2022jhk}, ALICE~\cite{wojnar2024applyinggenerativeneuralnetworks}, CMS~\cite{Vaselli:2858890,CMS-DP-2025-016,Dreyer:2024bhs,Dreyer:2025zhp}, LHCb~\cite{barbetti2024lamarrlhcbultrafastsimulation} and EIC~\cite{giroux2025generativemodelsfastsimulation} detectors.

Neural network-based surrogate models are naturally compatible with and accelerated by Graphical Processing Units (GPUs).  However, GPUs are energy intensive and are typically most efficient at large batch sizes, while event generation tends to operate in the batch-size-one regime, e.g. calorimeter showers are generated one at a time and not all at once across an event(s).
Field-programmable gate arrays (FPGAs) offer a complementary option in heterogeneous computing environments, providing low-latency and power-efficient inference suited to real-time or high-throughput use cases. 
FPGAs are already incorporated into the trigger and data-acquisition systems of the ATLAS and CMS detectors, making them a readily available resource during shutdown periods~\cite{Aad_2024,Hayrapetyan_2024}.
Existing studies have primarily considered FPGAs for online ML inference, benefiting from their deterministic latency and the reuse of signal-processing hardware. However, recent advances in modern FPGA devices, particularly their increasing logic density, also make them attractive for offline inference. In this work, we propose fast simulation as a testbed for exploring FPGA-based offline applications and for evaluating FPGAs as an heterogeneous computing element.
The ability to run generative simulation algorithms on FPGAs could lower latency and power of MC production while fully leveraging existing computational resources.

A practical challenge is that state-of-the-art generative models, such as normalizing-flow-based approaches~\cite{caloflow2,caloinn,l2lflow},
diffusion-based methods~\cite{caloscore2,calodiff},
and conditional-flow-matching models~\cite{calodream},
often require large networks together with complex architectures and operators to achieve the desired fidelity, making their deployment on resource-constrained hardware non-trivial~\footnote{Hybrid classical-quantum generative models are also beginning to be explored for generative simulation tasks ~\cite{hoque2024caloqvaesimulatinghighenergy}.}. Recent studies~\cite{BitHEP} exploring methods for compressing these models show promising reductions in size, although the resulting models remain still significantly larger than what is practical for realistic hardware implementation.
Further work on model compression techniques like pruning and quantization is therefore essential to minimize the algorithm's computational footprint so that it can fit within realistic FPGA systems.
These optimizations generally introduce some degradation in performance, but for many simulation use cases a modest loss in accuracy is an acceptable trade-off in exchange for substantially increased simulation throughput and reduced power consumption.

In this paper, we demonstrate the utility of FPGAs for fast and efficient ML-based calorimeter simulation. 
We make use of prior work from the 2022 CaloChallenge~\cite{Krause_2025}, where a variety of generative models were implemented ranging from variational autoencoders~\cite{Kingma:2013vae} to normalizing flows~\cite{Rezende:2015flows} and diffusion models~\cite{Ho:2020ddpm}. 
Our demonstration uses a compressed variational autoencoder model that can generate simulations faster than a GPU deployment at small batch sizes for a modest compromise in fidelity. 
This opens the door to the full exploitation of available FPGA resources at the LHC experiments, by using them for fast simulation during data-taking downtime and enables heterogeneous computing through streaming-like data transfer interface. This work also serves as an initial exploration of the use of FPGAs for offline tasks in high energy physics, showing that FPGA platforms can provide low and deterministic latency, high-throughput data synthesis and processing for specific applications, with potential future application for reconstruction and data compression.

%% file: sections/methods.tex
\section{Methods}
\label{sec:methods}

\subsection{Dataset}

We use the Calorimeter Simulation Challenge (CaloChallenge) datasets~\cite{Krause_2025} as the primary benchmark in this study. These datasets emulate key aspects of modern calorimeter systems and have become a standard testbed for R\&D on fast, machine-learning–based simulation, with well-characterized features and established state-of-the-art baselines. Importantly, they provide one of the first widely adopted, public benchmarks that is sufficiently challenging for a systematic evaluation of FPGA-oriented fast-simulation workflows.

\begin{figure}[t]
  \centering
  \includegraphics[width=0.5\textwidth]{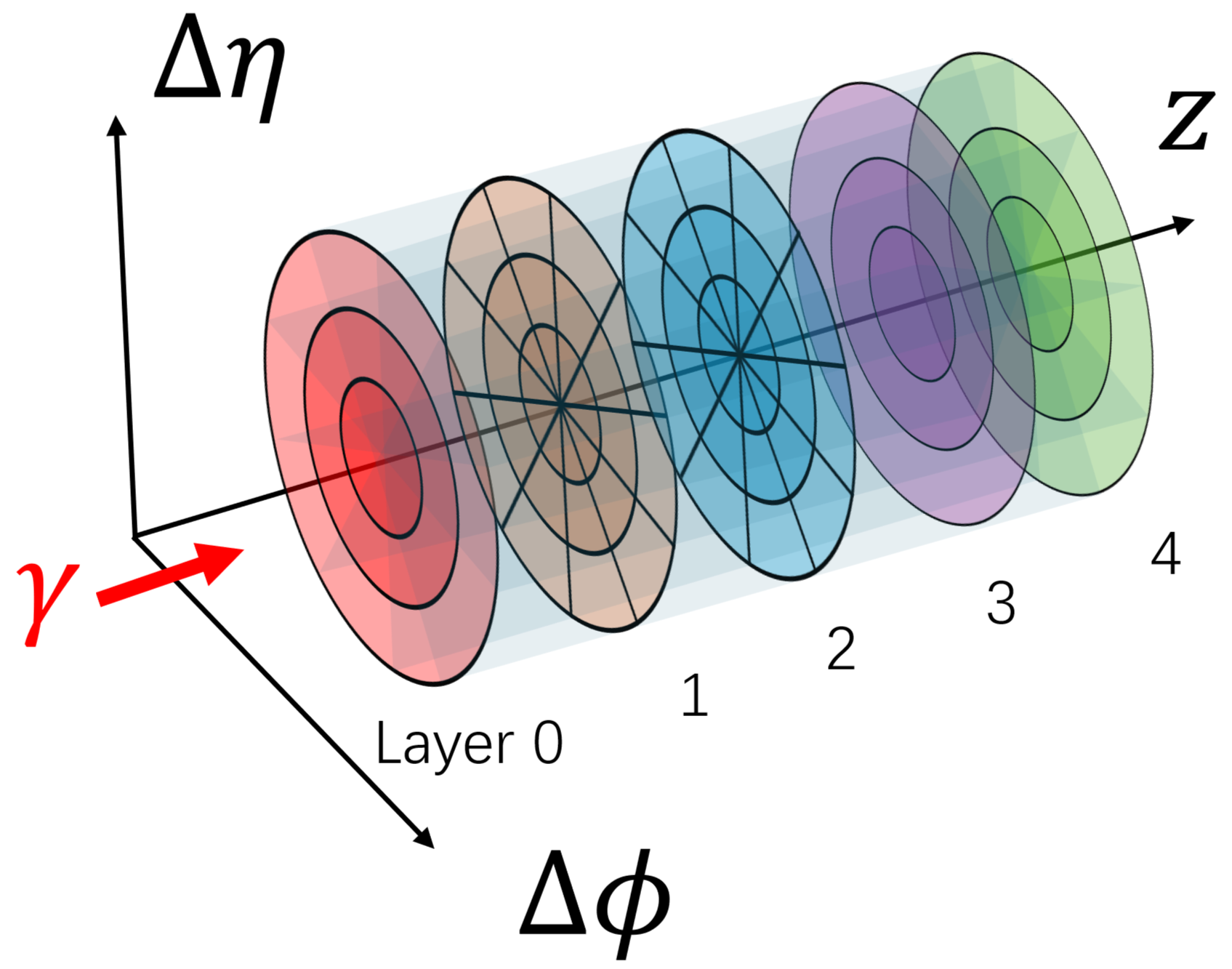}
  \caption{
    Demonstration of the dataset geometry. The variation of granularity along the radial direction for different layers is not reflected in the schematic for visualization reasons.
  }
  \label{fig:dataset_geometry}
\end{figure}

In this work we focus on the Photon Dataset 1~\cite{michele_faucci_giannelli_2023_8099322}, which contains 368-dimensional inputs and is derived from a prototype configuration in Ref.~\cite{ATLAS_OpenData_Calo} and studied in the context of the ATLAS experiment~\cite{ATL-SOFT-PUB-2020-006,Aad_2022}. 
Single photons are generated at the ATLAS calorimeter system surface and pointed toward the detector center, with $0.2<|\eta|<0.25$, corresponding to an oblique incidence on the calorimeter. 
The resulting detector response is simulated with the official ATLAS software chain based on \textsc{Geant4}~\cite{AGOSTINELLI2003250} using an idealized hit recording~\cite{ATLAS_OpenData_Calo}. 
Energy deposits are recorded in a five-layer geometry with irregular voxelization granularity $(8/160/190/5/5)$. The incident photon energy spans 256~MeV to 4~TeV, sampled at 15 logarithmically spaced discrete values. A schematic of the geometry is shown in Fig.~\ref{fig:dataset_geometry}. The dataset is also referred to as ``full-simulation,” in contrast to the ML-based fast simulation (``fast-simulation") studied in this work, and is used as the reference for physics performance.

\subsection{Preprocessing}

The data is preprocessed before feeding into the model. First, the 368 voxel energies undergo layer-wise normalization by dividing each voxel energy by the total energy of its respective layer. The resulting normalized voxel energy ratios $v_i$ are given by the following equation,
\begin{equation}
    v_{i} = E_i/ L_{l(i)},
\end{equation}
where $i$ is the index of each voxel, $E_i$ are the voxel energies, and $L_{l(i)} = \sum_{l(j)=l(i)} E_j$ is the total layer energy which is given by summing all of the voxel energies in layer $l(i)$.

In addition to providing the model with voxel energy ratios, the model also requires an energy response ratio and layer energy ratios to properly rescale the reconstructed voxel energy ratios. The energy response ratio is given by the following expression
\begin{equation}
    r = \frac{1}{\zeta}\frac{E_\mathrm{tot}}{E_\mathrm{inc}},
\end{equation}
where $r$ is the energy response ratio, $E_\mathrm{tot} = \sum_i E_i$ and $\zeta$ is a normalization factor\footnote{This factor is manually chosen so that 99.9 \% of the energy response ratios are between zero and one. A few high energy response events are intentionally left unnormalized so the distribution of energy responses has a mean of $\sim 0.5$.}. The layer energy ratios are the energy in a given layer divided by the total energy. The layer energy ratios $\ell_l$ are normalized by definition and given by
\begin{equation}
    \ell_l = L_l/E_\mathrm{tot},
\end{equation}
where $l$ is again the layer index.

The concatenation of voxel energy, energy response, and layer energy ratios is the model's input vector $x_k$ where the index $k$ goes from 1 to 374. 
In summary, for each training event, the model receives a 374 dimension input vector $x_k$ consisting of 368 voxel energy ratios $v_i$, 1 energy response ratio $r$, and 5 layer energy ratios $\ell_l$. 

In addition, the model also receives a conditional input. 
The model's conditional input is given by logarithmically scaling and then normalizing the incident energy. This transformation is represented by the following expression,
\begin{equation}
    x_\mathrm{con} = \log_2(E_\mathrm{inc})/\log_2(E_\mathrm{inc}^\mathrm{max}),
\end{equation}
where $x_\mathrm{con}$ is the conditional input, $E_\mathrm{inc}$ is the incident energy, and $E_\mathrm{inc}^\mathrm{max}$ is the maximum incident energy with a numerical value of $2^{22}$ MeV. This rescaling of the incident energies results in the network making improved distinctions between lower energy events.

\subsection{Generative Model}

We employ a conditional variational autoencoder (cVAE) model \cite{kingma2022autoencodingvariationalbayes,rezende2014stochasticbackpropagationapproximateinference,NIPS2015_8d55a249}, implemented with fully connected (dense) layers and an architecture inspired by the reference \texttt{DNNCaloSim} model~\cite{Krause_2025} from the CaloChallenge developed for other datasets. The DNN-based model is readily scalable in both size and precision and is well suited for resource-constrained hardware. Its regular structure maps efficiently to FPGA implementations.
It is comprised of an encoder network $q(z|x;x_\mathrm{con})$ that parametrizes the approximate posterior distribution of the latent variable $z$ given the input $x$, and a decoder network $p(x|z;x_\mathrm{con})$ that parametrizes the conditional likelihood of the data and reconstructs $x$ from samples of $z$. 
The encoder and decoder networks are trained by maximizing the evidence lower bound (ELBO) on the conditional log-likelihood,


\begin{equation}
    \log p(x|x_\mathrm{con}) \geq \mathbb{E}_{z\sim  q(z|x;x_\mathrm{con})}[\log p(x|z;x_\mathrm{con})] - D_{KL}(q(z|x;x_\mathrm{con})||p(z))\,.
\end{equation}
The right-hand side defines the ELBO: the first term is the expected conditional reconstruction log-likelihood under samples $z\sim q(z|x;x_\mathrm{con})$, and the second term is the Kullback-Leibler (KL) divergence~\cite{kullback1951information} which regularizes $q(z|x;x_\mathrm{con})$ toward the prior $p(z)$.
In practice, we minimize the negative ELBO by implementing the reconstruction term with a weighted binary cross-entropy on the decoder outputs $\tilde{x}$, and evaluating the KL divergence in closed form for a diagonal-Gaussian posterior with an unconditional standard normal prior $p(z)=\mathcal{N}(0,I)$. The training loss becomes the following,
\begin{equation}
    \mathcal{L} = - w_\mathrm{reco}\sum_k^{374} \left(  x_k \log \tilde{x}_k + (1-x_k)\log(1-\tilde{x}_k)  \right) +
    \frac{1}{2}\sum_i^{d_z} \left( \mu_i^2+\sigma^2_i - 1 - \log \sigma_i^2   \right),
\end{equation}
where $q(z| x,x_\mathrm{con})=\mathcal{N}(z|\mu,\mathrm{diag}(\sigma^2))$ is a diagonal-covariance Gaussian with mean 
$\mu$ and variance $\sigma^2$ predicted by the encoder, $d_z$ denotes the latent-space dimension, and $w_{\mathrm{reco}}$\footnote{$w_\mathrm{reco}$ has a numerical value of 374.} controls the relative weight of the reconstruction term.

The target vector $x\in \mathbb{R}^{374}$ is composed of per-layer voxel energy ratios, layer energy fractions, and an overall energy response ratio. The conditional encoder network has an input dimension of 375 ($x$ and $x_\mathrm{con}$)  and is followed by four dense layers with descending dimensionality. Each dense layer in the encoder is followed by a batch normalization layer and uses a leaky ReLU activation.
The decoder network begins with four dense layers in increasing dimensionality, mirroring the encoder. As in the encoder, each of these dense layers is followed by a batch normalization layer and use leaky ReLU activations. Following the fourth dense layer is another dense layer of dimension 374 which itself branches off into seven separate dense layers.
Five of these branching layers correspond to the five layers of the detector, with output dimensions of (8, 160, 190, 5, 5) respectively, and use softmax activations to reconstruct the 368 voxel energy ratios $\tilde{v}_i$ and enforce normalization within each calorimeter layer. The two remaining layers reconstruct the 5 layer energy ratios $\tilde{\ell}_l$ and energy response ratio $\tilde{r}$ with softmax and sigmoid activations respectively. The outputs are then concatenated into the final reconstructed output vector $\tilde{x}$. 

During generation we sample $z\sim p(z)=\mathcal{N}(0,I)$ and pass it along with $x_\mathrm{con}$ into the decoder to output $\tilde{x}$. For posterior sampling (used during training and for reconstructions), latent samples are instead drawn from $q(z| x,x_\mathrm{con})$ via the reparameterization $z=\mu+\sigma\odot\epsilon$, with $\epsilon\sim\mathcal{N}(0,I)$ (where $\odot$ represents element-wise multiplication).

\begin{figure}[tbh]
\centering
\includegraphics[width=0.8\textwidth]{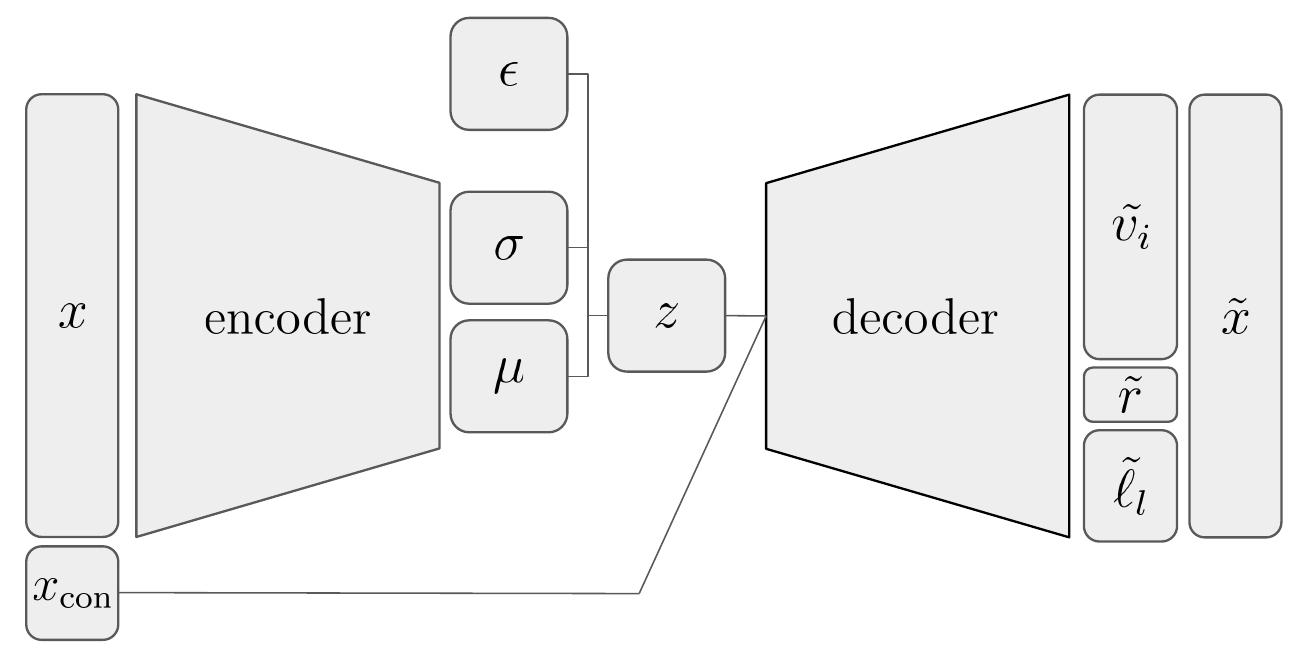}
\caption{Visualization of the VAE model. Forward propagation runs from left to right starting with the preprocessed input vector $x$ and condition $x_\mathrm{con}$ which feed into the encoder to output the vectors $\mu$ and $\sigma$. 
The condition $x_\mathrm{con}$ and the latent vector $z$ pass through the decoder and output the reconstructed ratios which are concatenated into the output vector $\tilde{x}$. 
\label{fig:model sketch} }
\end{figure}

For training, we adopt an eight-stage schedule, based on the reference \texttt{DNNCaloSim} model, where the learning rate is progressively decreased, and the batch size is adjusted, as summarized in Table~\ref{tab:train}. During each stage of training, 85\% of the data is randomly selected for training while the remaining 15\% is used for validation. The VAE is then optimized with the respective set of hyperparameters until the validation loss fails to improve for 10 epochs. After this early stopping, the model parameters are saved and used as the starting parameters of the next stage of training. 
The VAE was implemented in Keras 2.12~\cite{chollet2015keras} and QKeras 0.9~\cite{Coelho_2021} using TensorFlow 2.12~\cite{abadi2015tensorflow} as the backend, and trained with mini-batch gradient descent using the Adam optimizer~\cite{kingma2014adam}. Training was performed on an NVIDIA A100-SXM4-40GB GPU with 6,912 CUDA cores each clocked at 1410 MHz which resulted in a typical training time of 3 hours.

\begin{table}[t]
\centering
\begin{tabular}{ccccccccc}
\hline
training parameter &  1 &  2 &  3 &  4 &  5 &  6 &  7 &  8 \\
\hline
batch size & 100 & 50 & 100 & 25 & 100 & 50 & 25 & 100 \\
learning rate & $10^{-2}$ & $10^{-3}$ & $10^{-3}$ & $10^{-4}$ & $10^{-5}$ & $10^{-6}$ & $10^{-7}$ & $10^{-8}$ \\

\hline
\end{tabular}
\caption{Hyperparameters schedule for 8 stage training}
\label{tab:train}
\end{table}

\subsection{Decoder Codesign and FPGA Implementation}

The inference latency of the generative model is a primary consideration for fast calorimeter simulation. In the existing detector simulation stack, the conditioning information, such as the truth energy of the particle from upstream stages like tracker simulation, usually arrives sequentially. The calorimeter simulation operating in mini-batches, or in particular with batch-size-one generation, is essential for seamless integration into the existing simulation chain, as it avoids the need for additional buffering or complex service scheduling. This requirement guides the design of the model and its implementation on the FPGA device. It also benefits from the developed low-latency I/O interfaces designed for streaming packets~\cite{10.1145/3731599.3767569}, enabling heterogeneous computing architectures for the simulation task.

Only the decoder part of the VAE architecture is required during generative inference, which reduces the hardware resource requirements. However, implementing this decoder-only model on a low-latency platform such as an FPGA still requires substantial compression, including quantization, pruning, and dimensionality reduction, in order to fit within the available resources.


To evaluate the effect of these compression techniques with respect to latency and fidelity, we study two model versions, VAE-GPU and VAE-FPGA, which share the same model architecture and dimensionality as described in the previous subsection but with different precision and compression applied. VAE-GPU uses floating point precision and targets a GPU backend. It serves as a reference to contextualize the impact of the compression techniques. VAE-FPGA shares the same dimensionality as VAE-GPU, but its decoder is constructed from quantized dense layers and undergoes pruning to remove redundant neurons and synapses. Pruning is applied to the decoder during training using a constant-sparsity schedule. The schedule begins at training step 2000, allowing the network to first reach a stable initialization, and is updated every 100 steps thereafter. At each update, a fixed fraction of the smallest-magnitude weights is zeroed in order to maintain a target sparsity of 85\% (the fraction of zero-valued weights), effectively reducing the number of  parameters and making the network ``sparse''. This sparsity translates directly into hardware savings during synthesis, as~\hlsforml~refrains from instantiating multiply-accumulate units for zero-valued weights, reducing the number of look-up tables (LUTs) and digital signal processors (DSPs) allocated.

Nearly all VAE-FPGA dense layers use \texttt{ap\_fixed<6,2>} for weights and 
\texttt{ap\_fixed<8,3>} for biases, with two exceptions. The dense layer preceding 
the softmax activation that approximates the layer energy ratios uses 
\texttt{ap\_fixed<8,3>} for weights and \texttt{ap\_fixed<10,3>} for biases, 
and the dense layer preceding the sigmoid activation that approximates the energy 
response ratio retains full 32-bit floating-point precision, as accurate energy 
response estimation is critical for faithful shower generation. This layer is eventually processed with post-training quantization as will be discussed at the end of this section.
A quantitative summary of each model version's dimensionality and compression hyperparameters is shown in Table \ref{tab:dimensions}. The reference DNN model is also included for completeness; however, it was trained on a different dataset and is therefore provided only for indicative comparison. To further contextualize performance, we also report results from two other benchmarks.
The first is the current fastest GPU-based model in the CaloChallenge community, \texttt{CaloVQ}~\cite{liu2024calovqvectorquantizedtwostagegenerative}, which serves as a well-tuned benchmark for GPU platforms.
The second is the fastest model for a batch size of 1, namely \texttt{CaloINN}~\cite{caloinn}, which provides a GPU benchmark for small batch sizes where FPGA implementation is hypothesized to excel. 

The VAE-FPGA decoder is synthesized for FPGA implementation using ~\hlsforml~\cite{fastml_hls4ml, Duarte:2018ite}. 
For synthesis we utilize the ``resource" strategy and \texttt{io\_stream} as the IO\_type. All quantized dense layers use a  reuse factor equal to their input dimension. This relatively high reuse-factor scheme reduces resource utilization and simplifies HLS scheduling and operator binding. The default internal precision is \texttt{ap\_fixed$<$16,6$>$} while each layer type uses a specific precision, as described in Table \ref{tab:hls4ml_layer_precisions} of Appendix~\ref{app:hls4ml}. Notably, the dense layer and subsequent sigmoid activation (which approximate the energy response ratio) were specifically trained with full floating-point precision and post-quantized by~\hlsforml~for FPGA implementation, in contrast to all other layers and activations, which underwent quantization-aware training using QKeras 0.9 prior to the ~\hlsforml~ conversion for higher compression ratios.
The target FPGA is an AMD Xilinx Virtex UltraScale+ (xcvu13p-flga2577-2-e).

\begin{table}[t]
\centering
\begin{tabular}{ccccc}
\hline
Model & Latent Dimension & Trainable Parameters & Sparsity & Precision  \\
\hline
\texttt{DNNCaloSim} & 50 & 3,169,663 & 0\%& FP32 \\
\texttt{CaloINN} & 50 & 18,821,350 & 0\%&  FP32 \\
\texttt{CaloVQ} & 242 & 2,152,637 & 0\%&  FP32 \\
VAE-GPU & 30 & 234,884 & 0\% & FP32 \\ 
VAE-FPGA & 30 & 234,884 & 85\% & QINT16 \\
\hline
\end{tabular}
\caption{Summary of VAE-GPU, VAE-FPGA, and reference models, comparing size and application of compression techniques. “FP32” denotes the model using 32-bit floating-point numbers (single precision). “QINT16” denotes a quantized 16-bit fixed-point representation. Sparsity quantifies the fraction of trainable parameters that are zero-valued. The detailed layer-wise precision settings are listed in Table~\ref{tab:hls4ml_layer_precisions}.}
\label{tab:dimensions}
\end{table}

%% file: sections/results.tex
\section{Results}
\label{sec:results}


The viability of the FPGA-based simulation implementation as compared to traditional CPU or GPU-based platforms is gauged by two performance metrics: fidelity, i.e. proximity of generated simulations to a Geant4 reference, and FPGA resources including generation latency. 
Results for the synthesized VAE-FPGA model are reported relative to the VAE-GPU model and to state-of-the-art methods from the literature, as summarized in Table~\ref{tab:dimensions}.

\subsection{Fidelity}

\begin{figure}[tbh]
\centering
\includegraphics[width=1.0\textwidth]{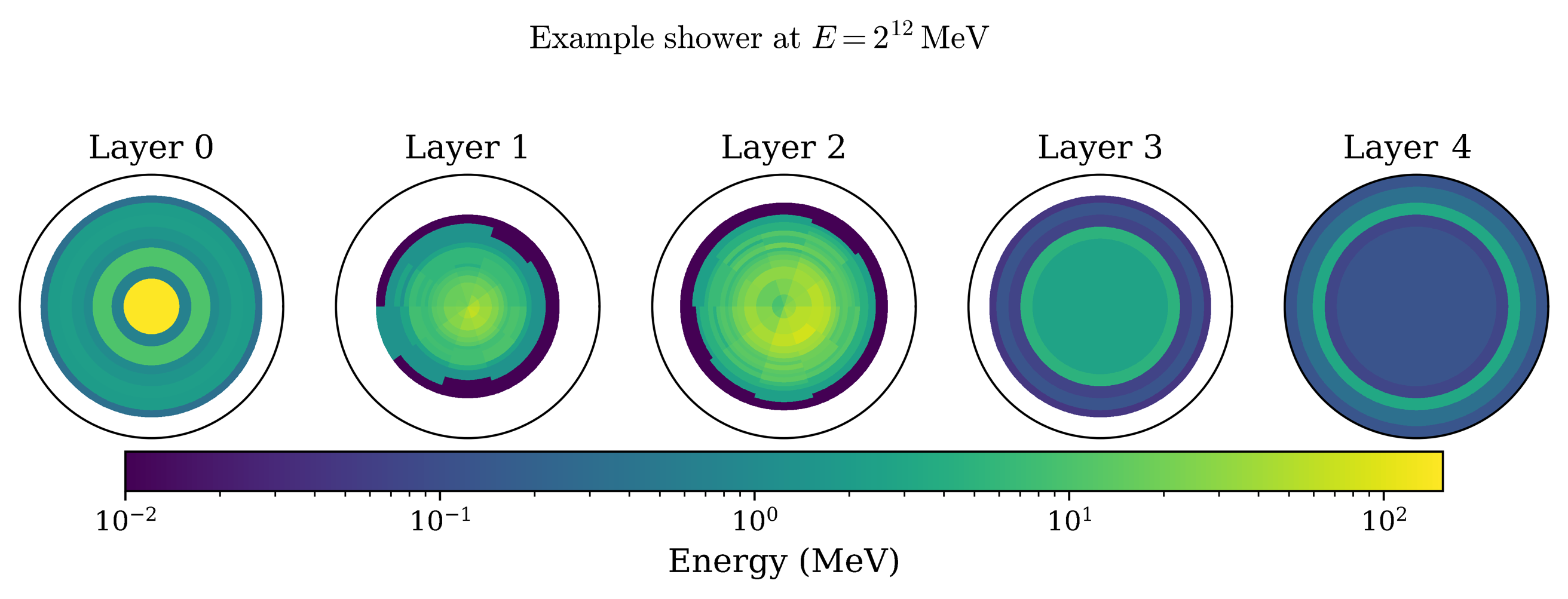}
\caption{Example of generated shower at $2^{12}$ MeV created by the VAE-FPGA model. \label{fig:showerEx} }
\end{figure}

Figure~\ref{fig:showerEx} shows an example generated shower using the synthesized VAE-FPGA model. 
The incident particle passes through 5 layers 
and deposits energy in each layer. The shower pattern shows that the model captures the expected spatial morphology and energy-deposition profile, including realistic lateral and longitudinal development, without unusual features or hot spots. This indicates physically consistent, statistically representative generation.

\begin{figure}[t]
  \centering
  \begin{subfigure}{\textwidth}
    \centering
    \includegraphics[width=1.0\textwidth]{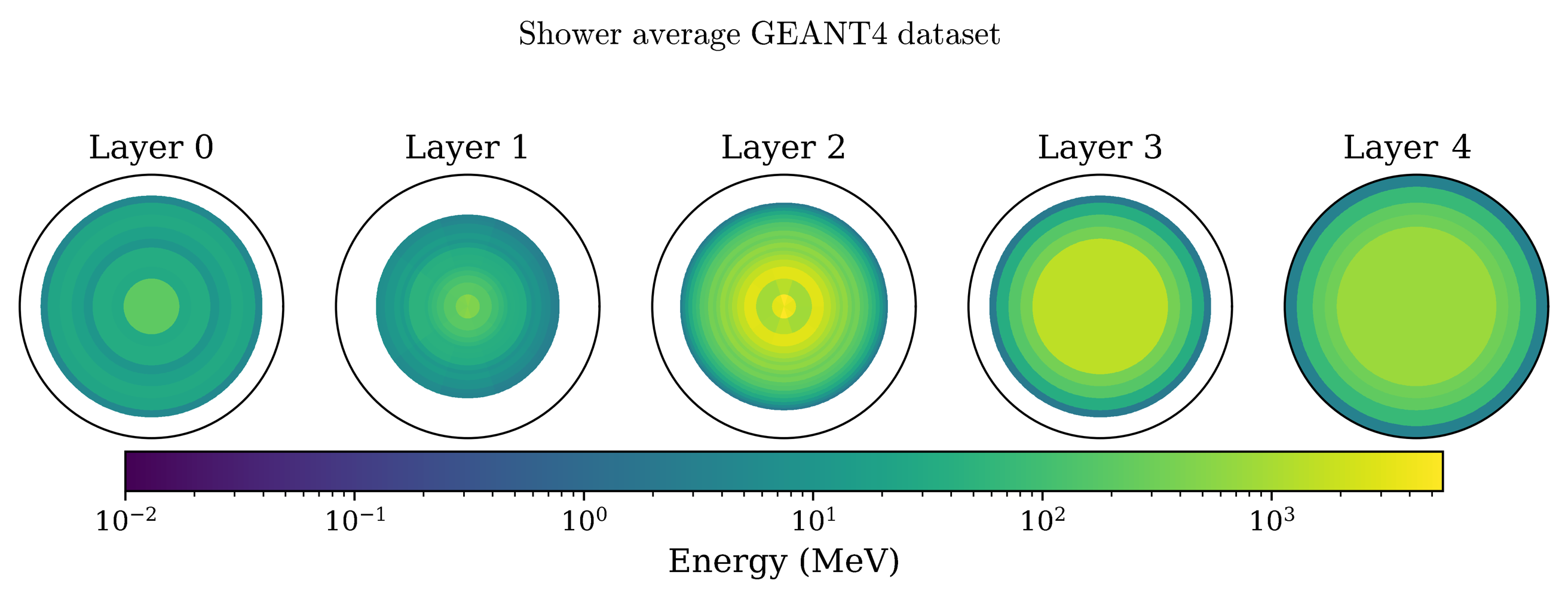}
    \label{fig:avg_ref}
  \end{subfigure}
  \vspace{-.12cm} 
  \begin{subfigure}{\textwidth}
    \centering
    \includegraphics[width=1.0\textwidth]{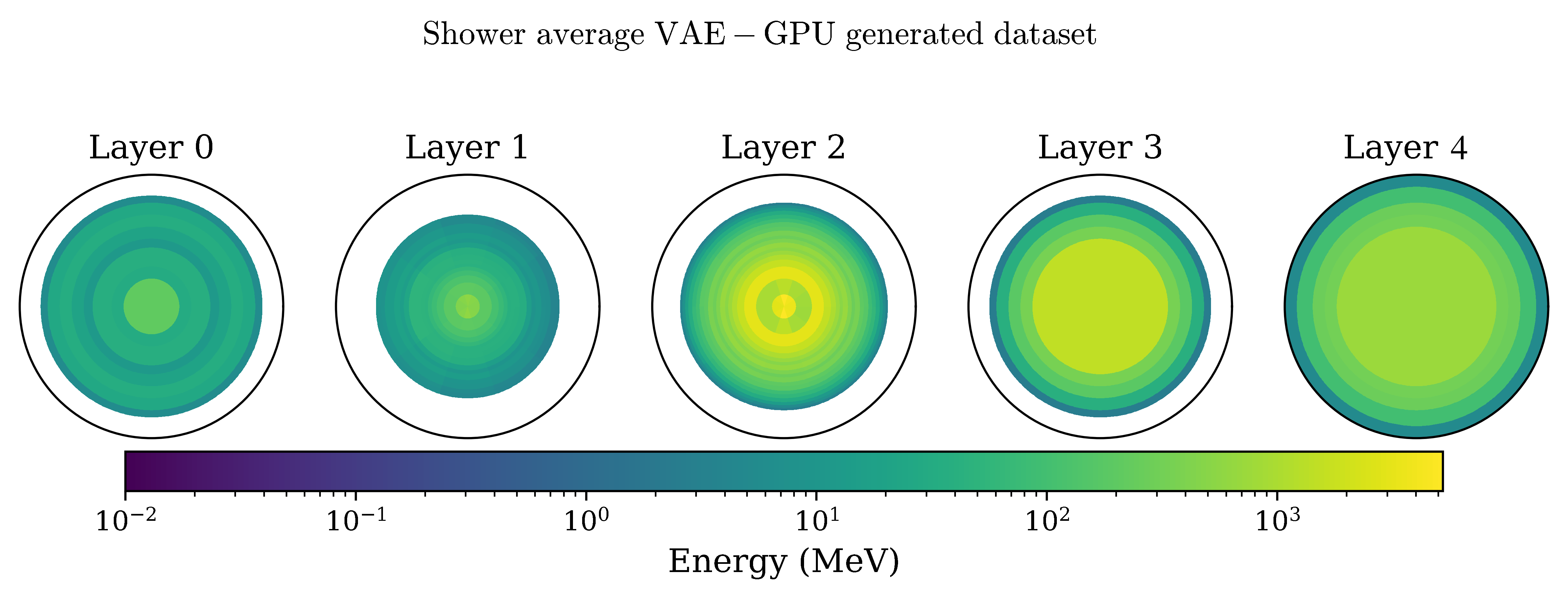}
    \label{fig:avg_gpu}
  \end{subfigure}
  \vspace{-.12cm}
  \begin{subfigure}{\textwidth}
    \centering
    \includegraphics[width=1.0\textwidth]{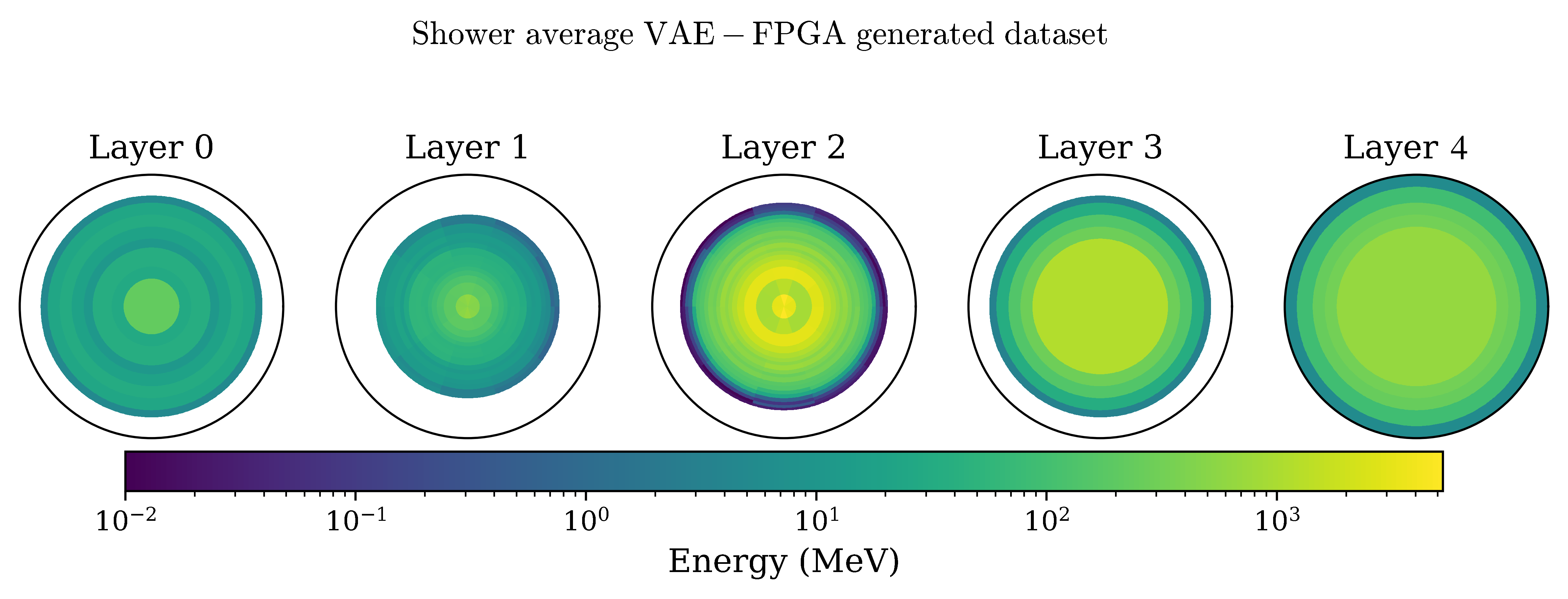}
    \label{fig:avg_gen}
  \end{subfigure}
  \caption{Average per-layer energy deposition, comparing the Geant4 truth (top), VAE-GPU generated (middle), and VAE-FPGA generated showers (bottom).}
  \label{fig:perlayer}
\end{figure}

Figure~\ref{fig:perlayer} shows the average per-layer energy deposition comparing 
the Geant4 truth to the VAE-GPU and VAE-FPGA generated results across all 5 layers. The VAE-GPU results are provided to illustrate the relative performance drop from model compression and FPGA implementation. The uniform and smooth angular distributions observed in layers 1 and 2 indicate that the generative model successfully captures rotational symmetry without explicit architectural enforcement, while remaining consistent with the training truth. This behavior reflects the underlying geometry and supports the feasibility of learning relevant physical shower features.

\begin{figure*}[t!]
    \centering
    \begin{subfigure}[t]{0.5\textwidth}
        \centering
        \includegraphics[width=0.8\textwidth]{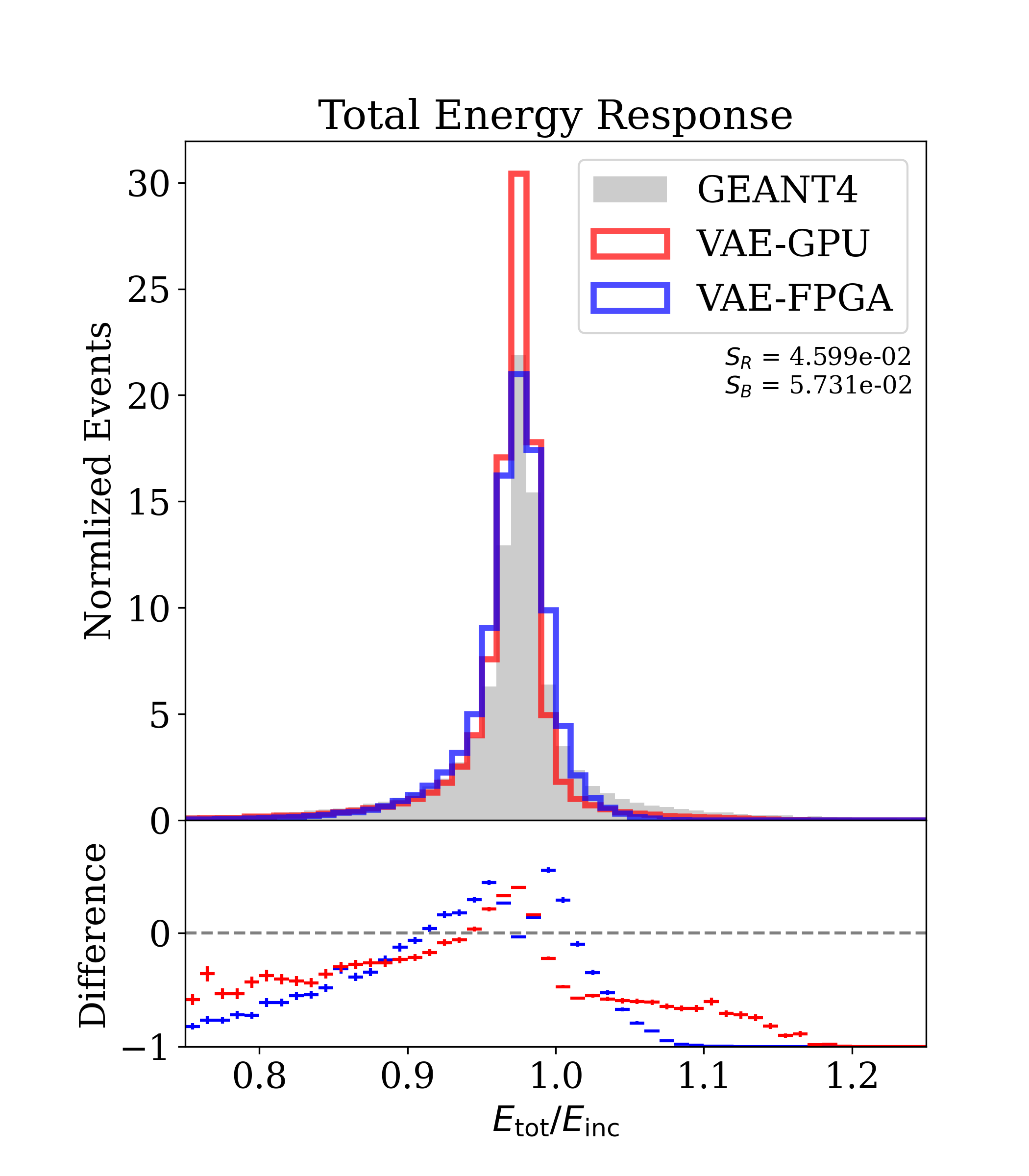}
    \end{subfigure}%
    \hfill
    \begin{subfigure}[t]{0.5\textwidth}
        \centering
        \includegraphics[width=0.8\textwidth]{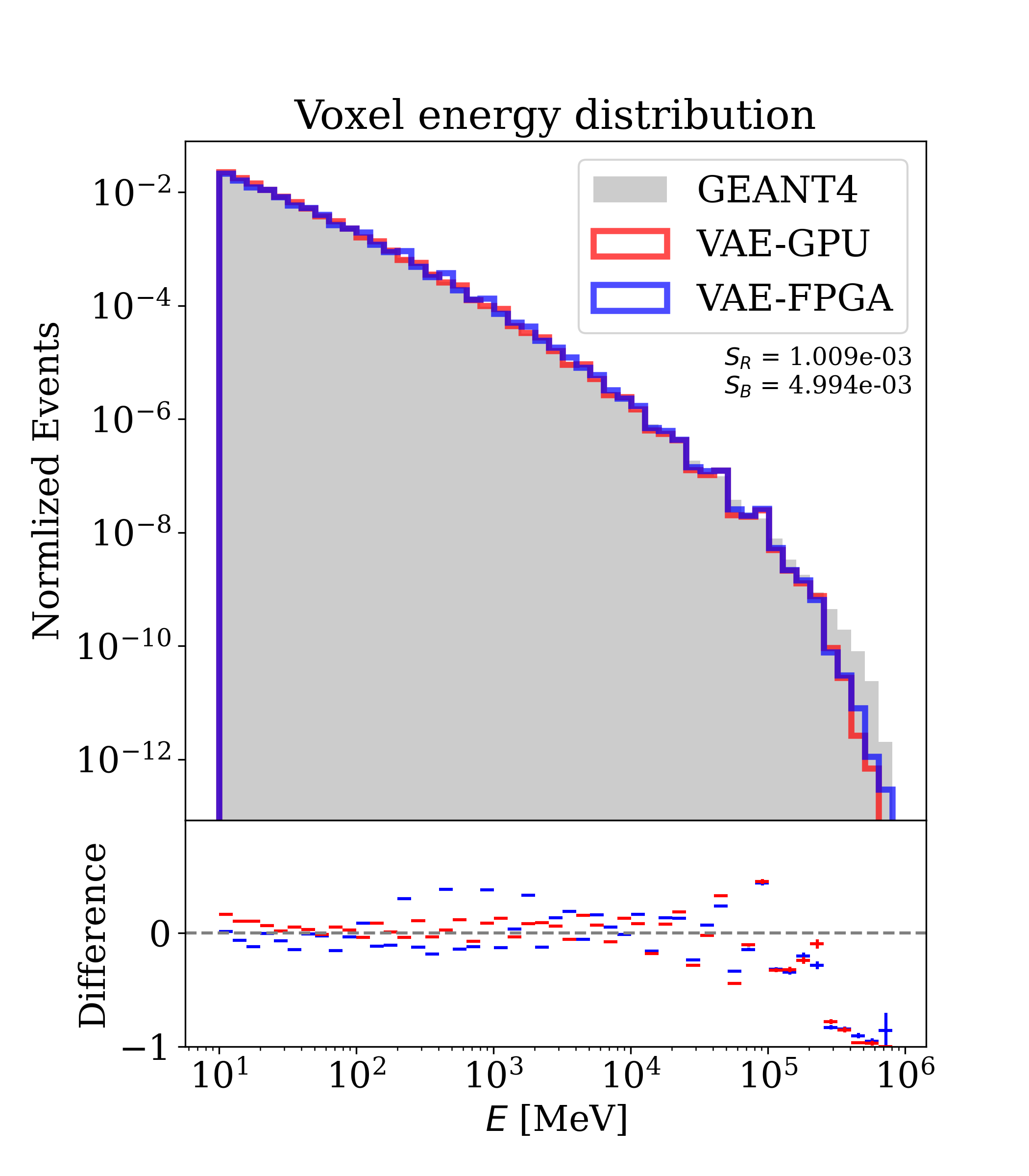}
    \end{subfigure}
    \caption{Energy response (left) and voxel energy distribution (right) histograms, comparing the Geant4 truth (gray), VAE-GPU generated shower (red), and VAE-FPGA generated shower (blue). The separation power for VAE-GPU $(S_R)$ and VAE-FPGA $(S_B)$ is provided for each feature. 
    \label{fig:energyresponse}}
\end{figure*}

Figure~\ref{fig:energyresponse} provides histograms of the resulting calorimeter energy deposition and overall energy distribution per voxel, comparing Geant4 truth to the VAE-GPU and VAE-FPGA generated result. 
The \textit{separation power S}, introduced in Eq.~\ref{eq:eva_1}, is employed to quantify the discrepancy between the reference full-simulation histogram ($h$) and the corresponding fast-simulation prediction ($h'$) at the same condition:

\begin{equation}
\mathrm{S} \equiv \sum_{i=1}^{N_{\mathrm{bin}}} \frac{(h'_i - h_i)^2}{2\,(h'_i + h_i)} \, .
\label{eq:eva_1}
\end{equation}

With good qualitative agreement of key features such as peak location and distribution and tail shapes, along with $S$ values below 0.1, these results further confirm good quality of VAE generated samples, even after compression for FPGA deployment. 

To further evaluate the fidelity of the fast simulation, a set of physics-motivated evaluation metrics is defined using one-dimensional histograms of key observables, following a compatible definition in Ref.~\cite{Krause_2025}. 
The set of evaluated observables comprises the energy deposited in individual calorimeter voxels ($E_i$), the total reconstructed energy response, and several shower-shape variables that are essential for accurate object reconstruction. 
A detailed definition of these observables is given in Table~\ref{tab:v}. The true incident photon energy, denoted by $E_{\mathrm{inc}}$, is used as a conditioning variable. 
The quantity $\Delta\eta_i(K)$ specifies the position (in millimeters) of voxel $i$ in the $\Delta\eta$ direction for calorimeter layer $K$, with an analogous definition for $\Delta\phi_i(K)$ in the $\Delta\phi$ direction. 

\begin{table}[t]
\centering
\begin{tabular}{ccc}
\hline
Symbol & Observable & Definition \\
\hline
$E_i$ & Voxel energy (index $i$) & Output of the fast simulation \\
$E_{\mathrm{tot}}$ & Total deposited energy & $\sum_i E_i$ \\
$E_{\mathrm{tot}}/E_{\mathrm{inc}}$ & Total energy response & $\sum_i E_i / E_{\mathrm{inc}}$ \\
$E(K)$ & Layer energy (layer $K$) & $\sum_{i \in K} E_i$ \\
$\overline{\eta}(K)$ & Energy centroid in $\Delta\eta$ (layer $K$) & $\sum_{i \in K} \Delta \eta_i\, E_i \,/\, E(K)$ \\
$\sigma_{\eta}(K)$ & Shower width in $\Delta\eta$ (layer $K$) & $\sqrt{\sum_{i \in K} (\overline{\eta}(K) - \Delta\eta_i)^2 \, E_i\,/\, E(K)}$ \\
$\overline{\phi}(K)$ & Energy centroid in $\Delta\phi$ (layer $K$) & $\sum_{i \in K} \Delta \phi_i\, E_i \,/\, E(K)$ \\
$\sigma_{\phi}(K)$ & Shower width in $\Delta\phi$ (layer $K$) & $\sqrt{\sum_{i \in K} (\overline{\phi}(K) - \Delta\phi_i)^2 \, E_i\,/\, E(K)}$ \\
\hline
\end{tabular}
\caption{Summary of physics observables used for the evaluation of simulation fidelity.}
\label{tab:v}
\end{table}

\begin{figure}[t]
  \centering
  \begin{subfigure}{0.33\textwidth}
    \centering
    \includegraphics[width=\linewidth]{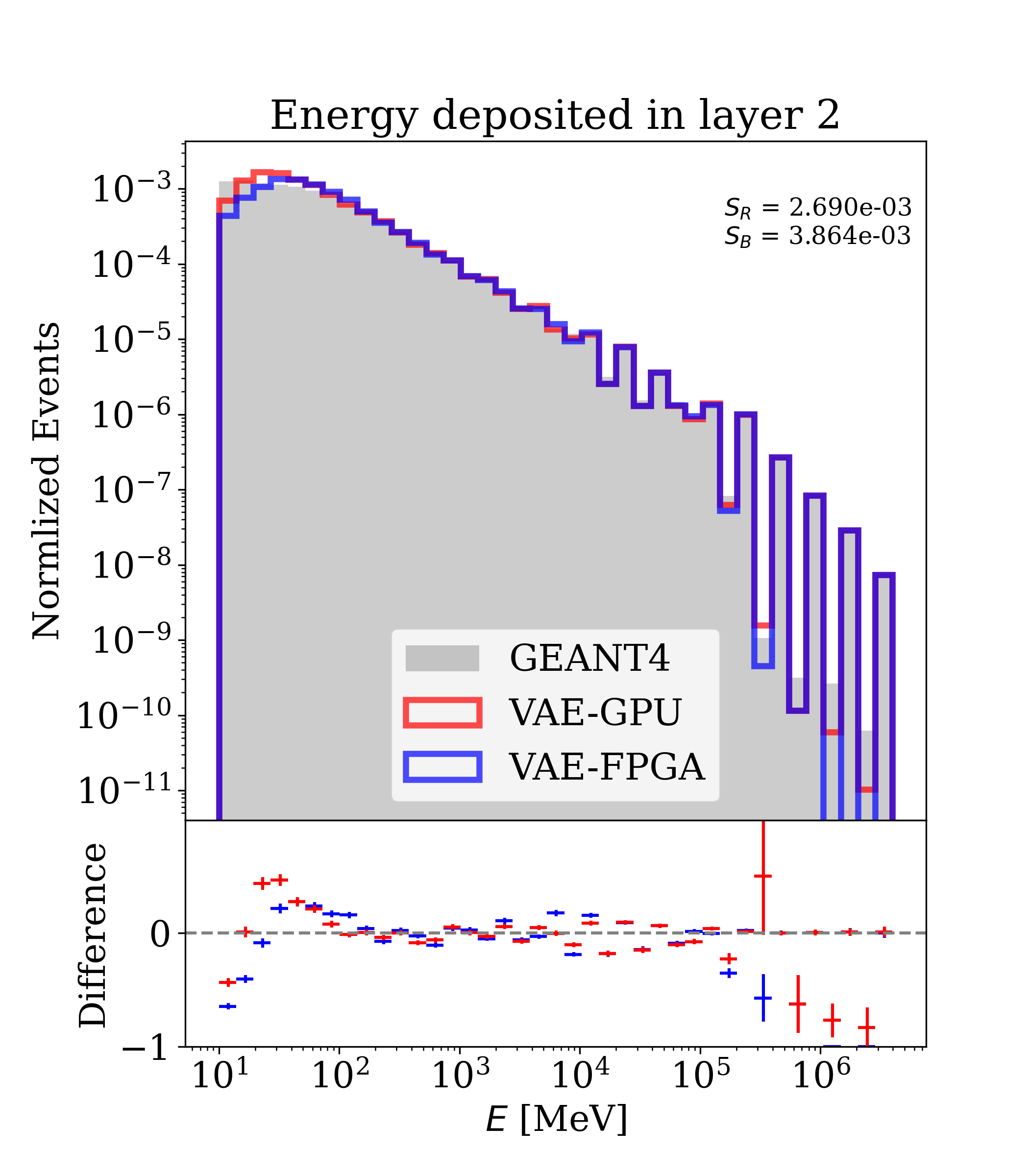}
    \label{fig:top1}
  \end{subfigure}\hfill
  \begin{subfigure}{0.33\textwidth}
    \centering
    \includegraphics[width=\linewidth]{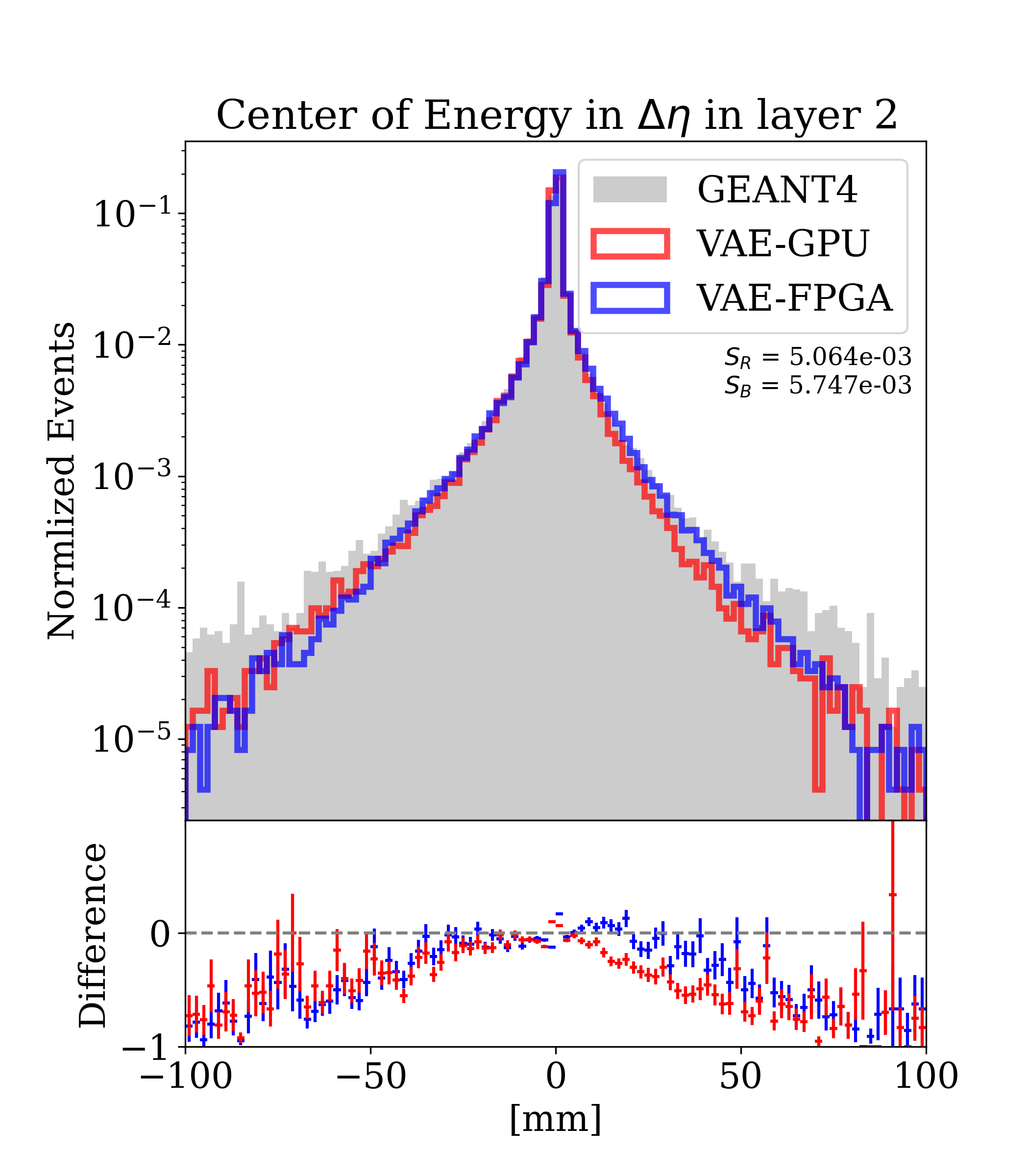}
    \label{fig:top2}
  \end{subfigure}\hfill
  \begin{subfigure}{0.33\textwidth}
    \centering
    \includegraphics[width=\linewidth]{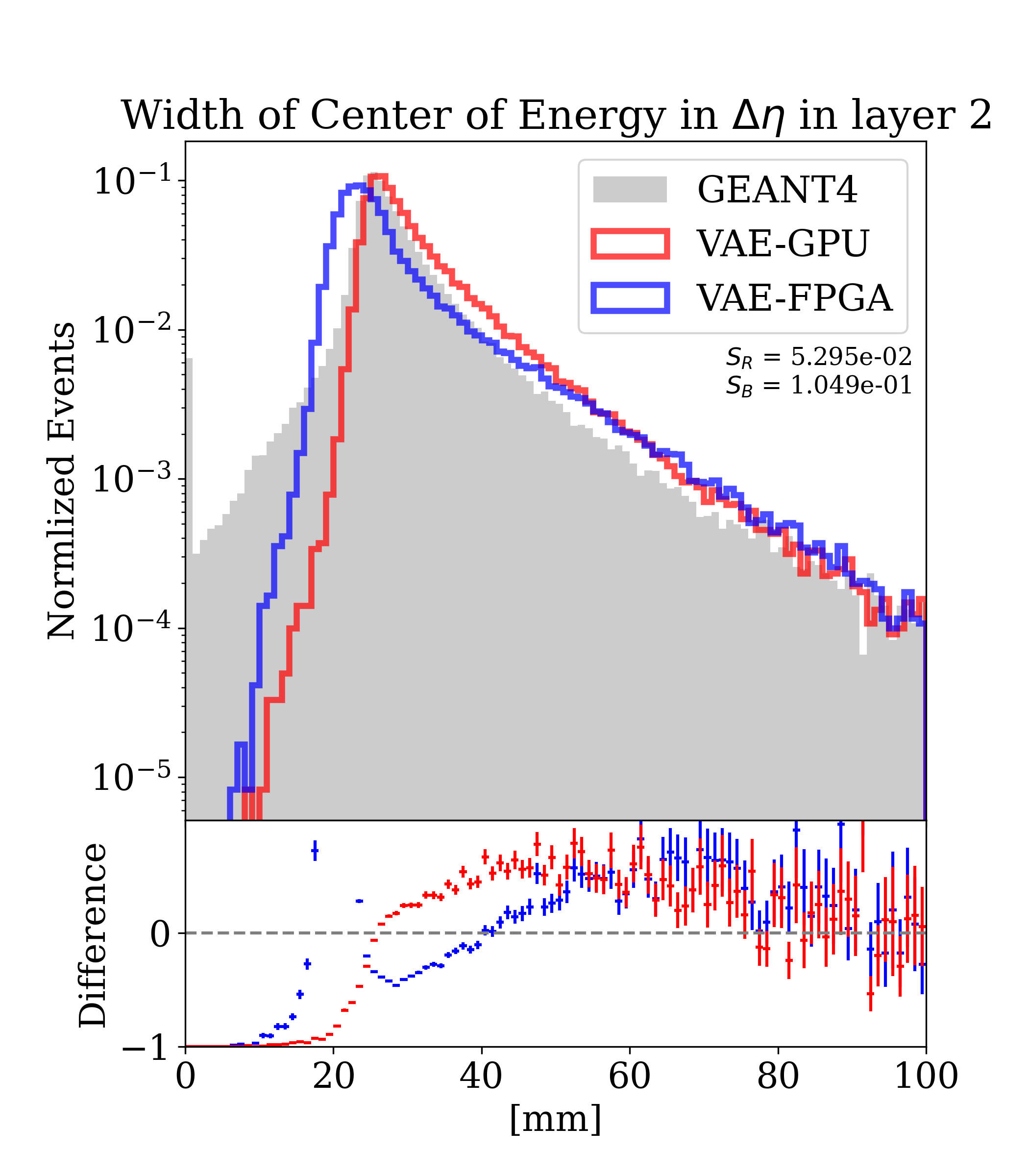}
    \label{fig:top3}
  \end{subfigure}

  \vspace{-1.8em} 

  \begin{subfigure}{0.33\textwidth}
    \centering
    \includegraphics[width=\linewidth]{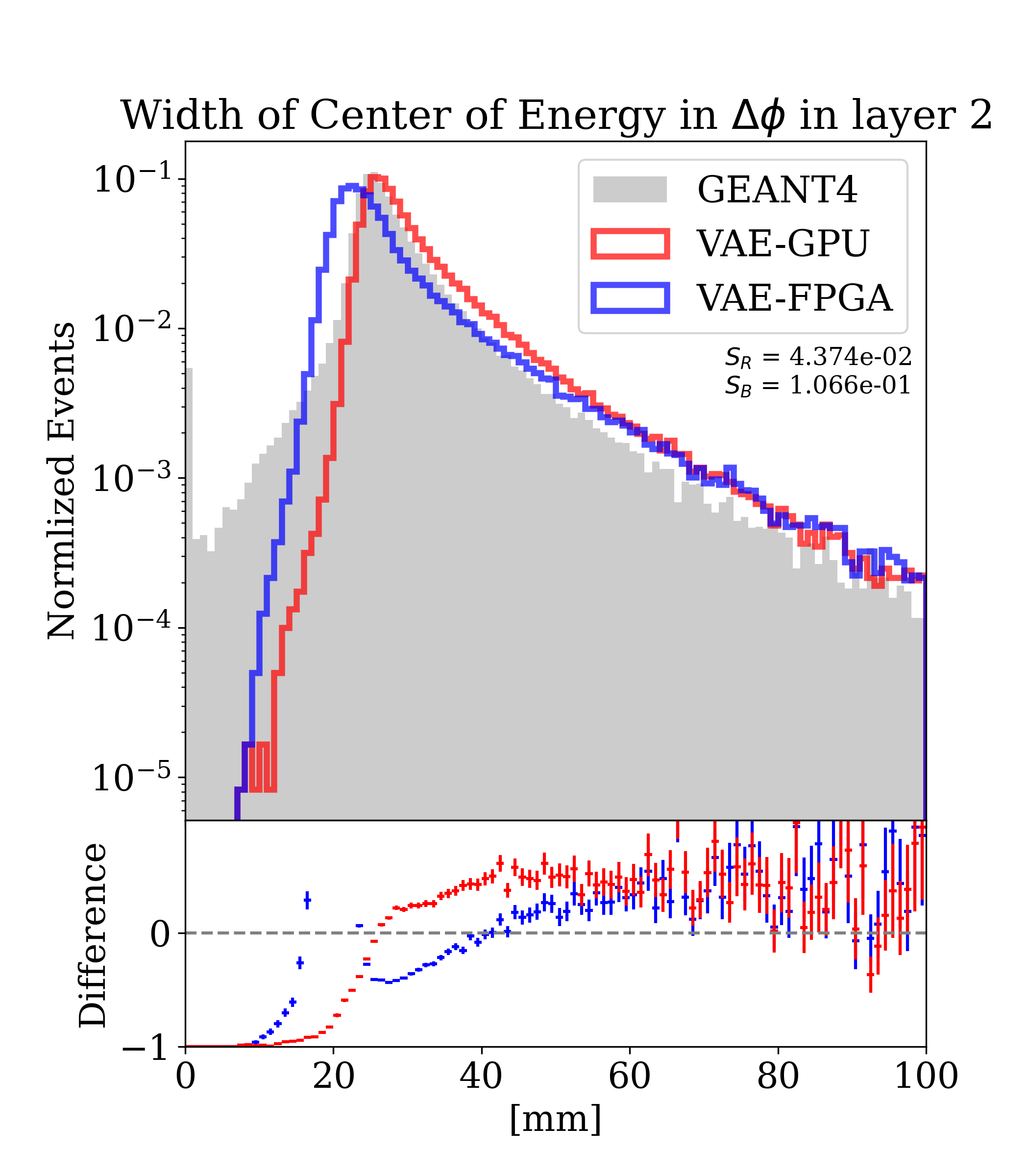}
    \label{fig:bot1}
  \end{subfigure}\hspace{0.03\textwidth}
  \begin{subfigure}{0.33\textwidth}
    \centering
    \includegraphics[width=\linewidth]{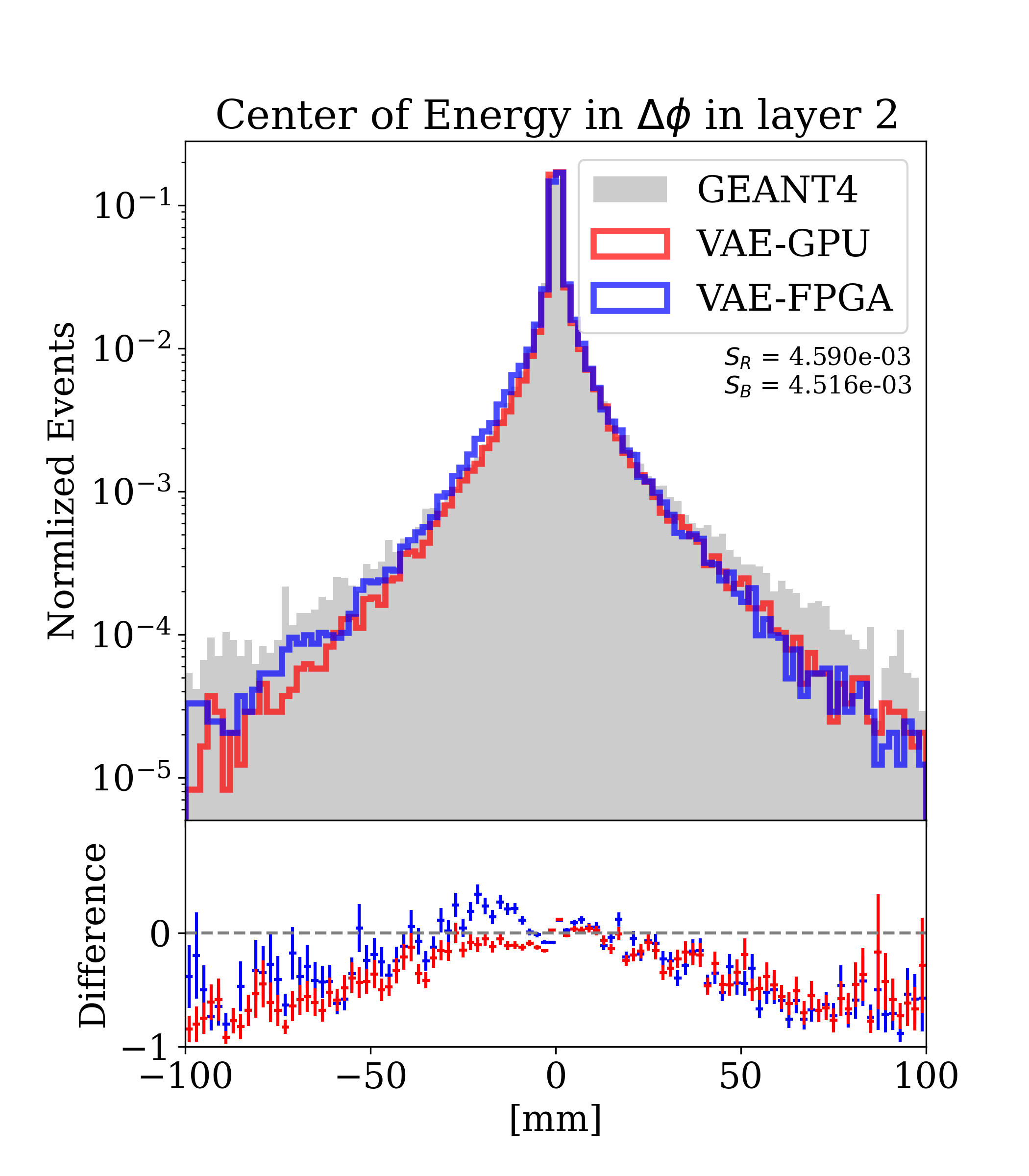}
    \label{fig:bot2}
  \end{subfigure}

  \caption{Distributions of physics variables comparing the Geant4 truth (gray), VAE-GPU generated shower (red), and VAE-FPGA generated shower (blue). Layer 2 is selected for demonstration. Specific histograms show layer energy (top left), shower center (top center) and width (top right) in $\Delta\eta$, shower center (bottom left) and width (bottom right) in $\Delta\phi$. The separation power for VAE-GPU $(S_R)$ and VAE-FPGA $(S_B)$ is provided for each feature. 
  \label{fig:layer2_phys_variables}}
\end{figure}

Figure~\ref{fig:layer2_phys_variables} shows the distributions of the shower-shape observables defined in Table~\ref{tab:v}, comparing the Geant4 truth, the VAE-GPU generated, and VAE-FPGA generated results for layer 2, which is selected for demonstration. The histograms cover the layer energy $E(K)$, the energy centroids $\overline{\eta}(K)$ and $\overline{\phi}(K)$, and the corresponding shower widths $\sigma_\eta(K)$ and $\sigma_\phi(K)$. These variables jointly characterize the lateral development of the shower and its spatial localization within the calorimeter. The observables and their correlations match the reference simulation closely near the mean, with moderate to high disagreement at small shower widths, as expected given the limited capacity of the VAE. However, the generated distributions of the VAE-FPGA model track those of the VAE-GPU within the level of agreement expected given the impact of quantization, pruning, and different training frameworks (QKeras 0.9 was used for the quantization-aware training of the VAE-FPGA, while Keras 2.12 was used for the VAE-GPU).

\begin{figure}[t]
  \centering
  \begin{subfigure}{0.33\textwidth}
    \centering
    \includegraphics[width=\linewidth]{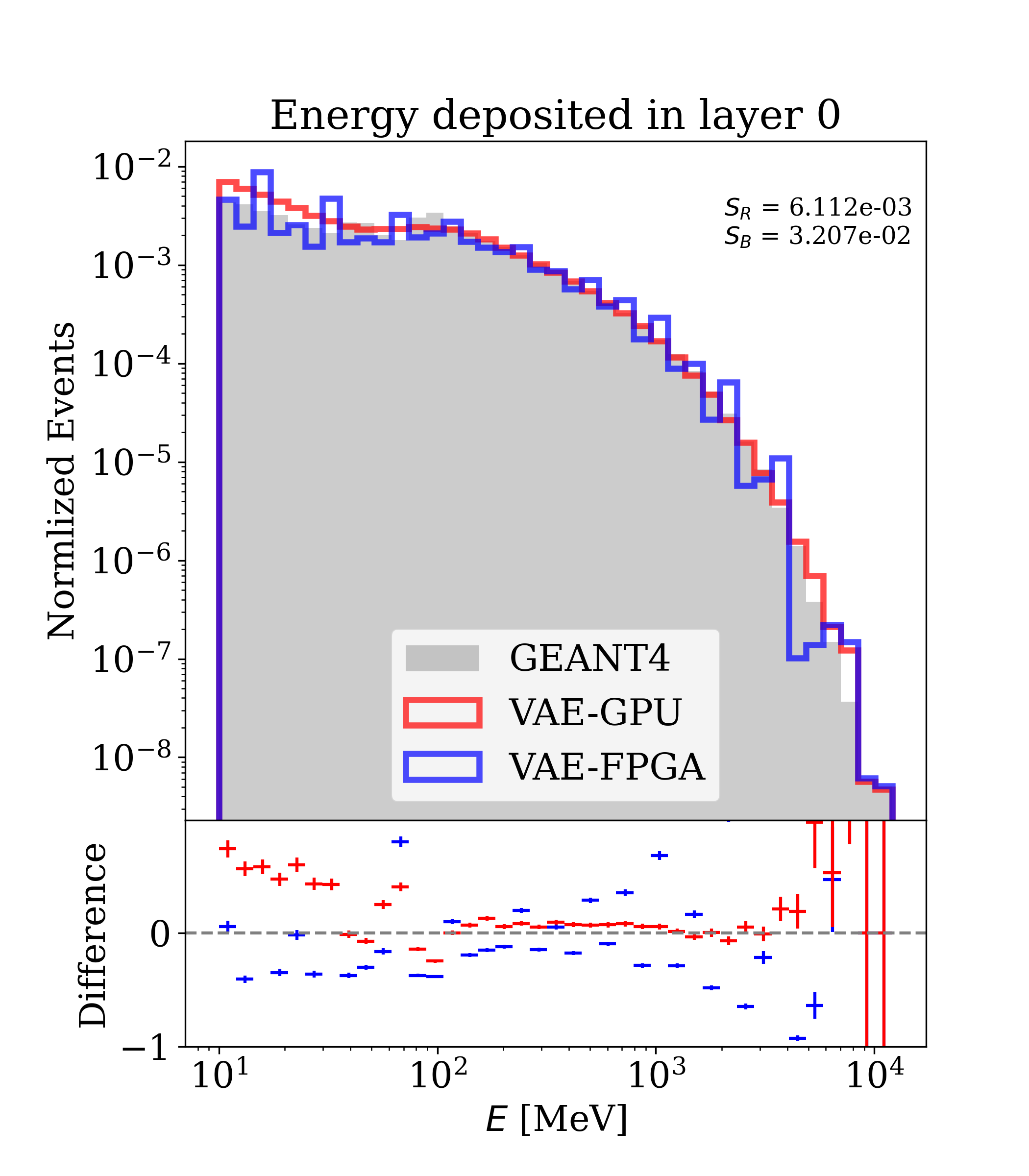}
  \end{subfigure}\hfill
  \begin{subfigure}{0.33\textwidth}
    \centering
    \includegraphics[width=\linewidth]{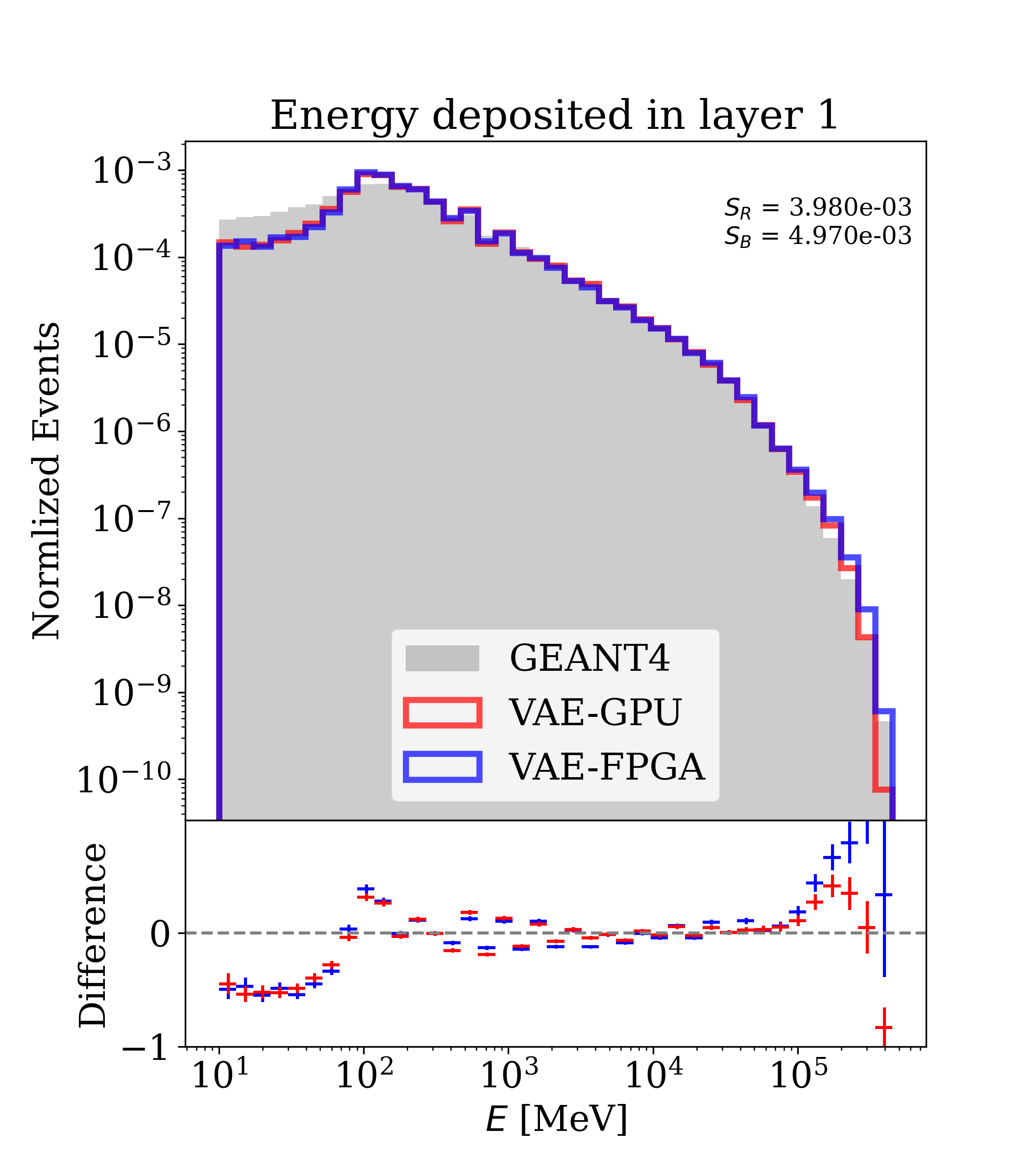}
  \end{subfigure}\hfill
  \begin{subfigure}{0.33\textwidth}
    \centering
    \includegraphics[width=\linewidth]{figures/E_layer_2_dataset_1-photons.png}
  \end{subfigure}

  \vspace{-0.5em} 

  \begin{subfigure}{0.33\textwidth}
    \centering
    \includegraphics[width=\linewidth]{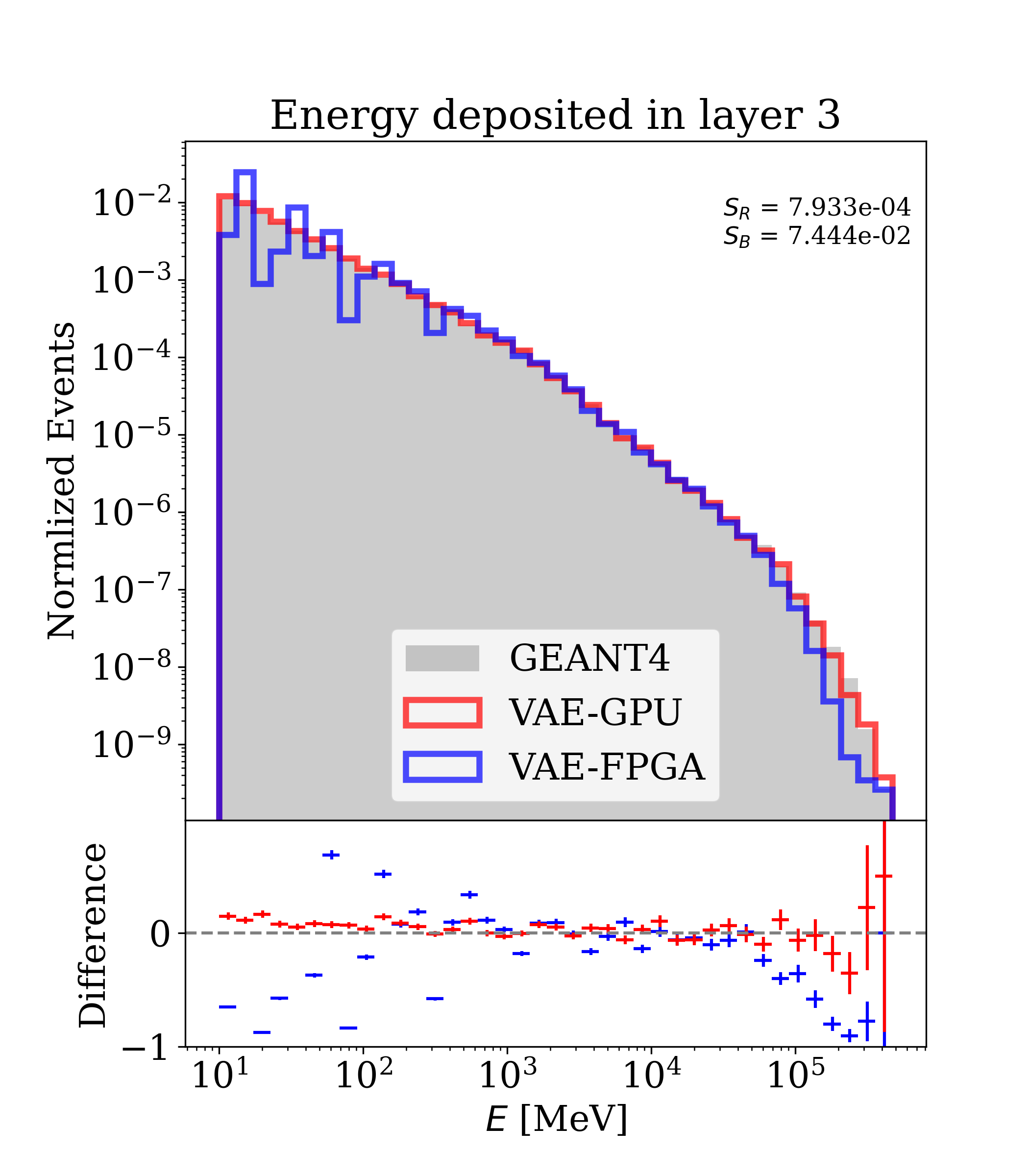}
  \end{subfigure}\hspace{0.03\textwidth}
  \begin{subfigure}{0.33\textwidth}
    \centering
    \includegraphics[width=\linewidth]{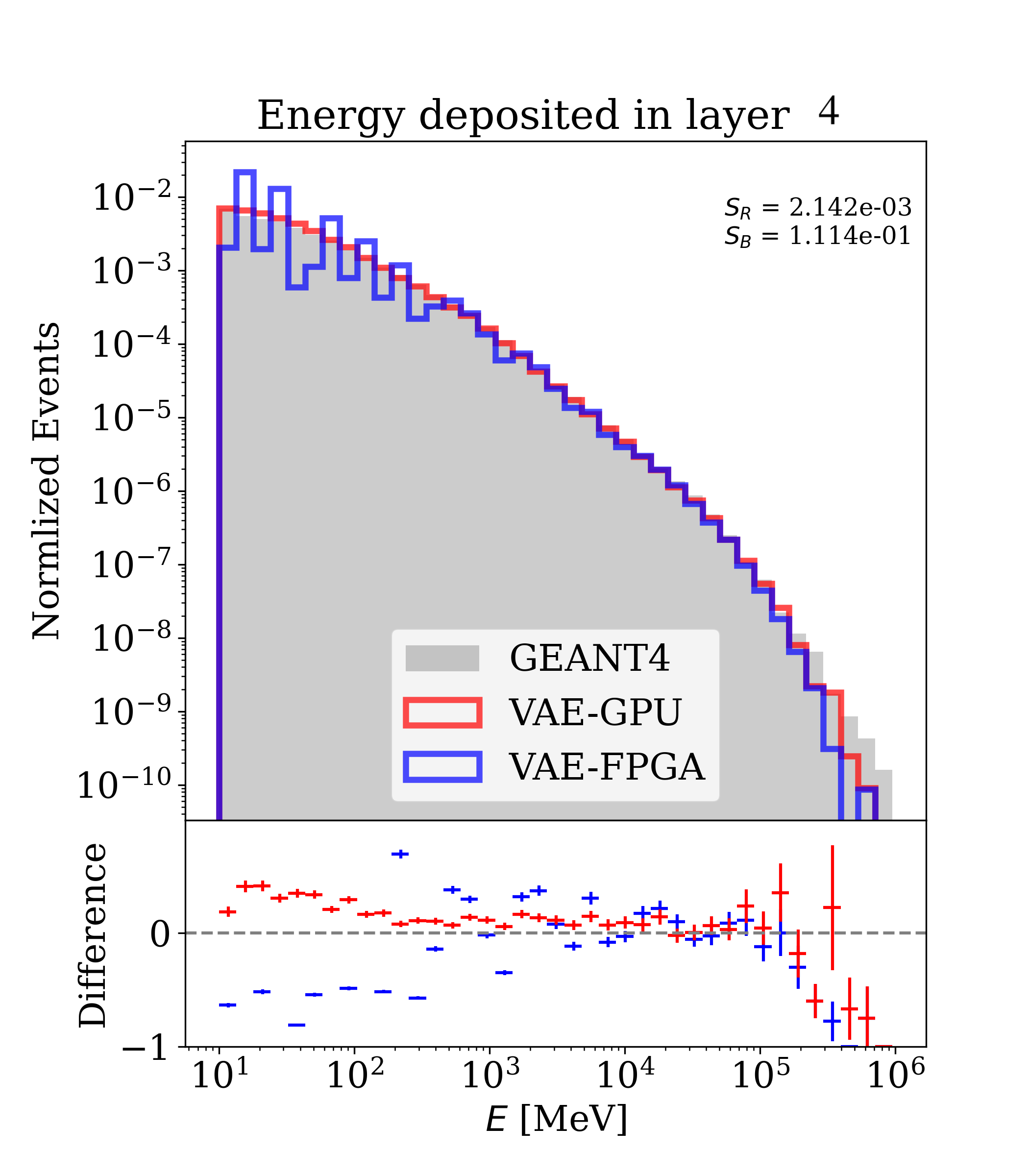}
  \end{subfigure}
  \caption{Per layer energy distribution from Geant4 truth (gray), VAE-GPU generated shower (red), and VAE-FPGA generated showers (blue). The separation power for VAE-GPU $(S_R)$ and VAE-FPGA $(S_B)$ is provided for each feature. }
  \label{fig:Ei_histo}
\end{figure}

Figure~\ref{fig:Ei_histo} shows the per-layer energy distributions comparing the Geant4 truth to the VAE-GPU and VAE-FPGA generated results. The energy distribution across layers along the $z$-direction captures the longitudinal development of the shower and serves as a key observable for benchmarking the generation accuracy, including the shower shape and voxel-energy correlations. The observables and their correlations closely match the reference simulation, with no visible artifacts, indicating that the generative model reproduces calorimeter information at the level of detailed shower shapes, beyond the total energy response.
Averaging over all physics variables gives an average separation of $S_\mathrm{avg}=0.054$ for the VAE-GPU and $S_\mathrm{avg} = 0.066 $ for the VAE-FPGA model, corresponding to a degradation of approximately 23\% after compression and hardware synthesis. 
Compared to truth, the most significant loss in performance arises from limited capacity of the VAE model architecture rather than the FPGA implementation itself.
While the simulation degradation is non-negligible, it indicates results of sufficiently high quality to consider partial offloading of generative simulation tasks to FPGAs, wherein tasks that require exceptionally high quality simulation can still rely on standard deployments. 

\subsection{Resources \& Latency}

Table~\ref{tab:results} shows the post-synthesis resource utilization and latency for the VAE-FPGA decoder, as reported by~\hlsforml. Resource usage is reported in terms of look-up tables (LUTs), flip flops (FFs), and digital signal processors (DSPs), with utilization expressed as the fraction of the target device's available resources. The FF and DSP utilization remain modest at 12\% and 15\% respectively, while LUTs are the dominant resource at 85\%. The combined resource usage is sufficient to deploy the full decoder generative model on a single modern commercial FPGA. However, LUT utilization is the dominant constraint, driven primarily by the large softmax activation layers, and is nearing capacity, limiting further improvements to the current VAE architecture.
In terms of speed, the current FPGA model outperforms all GPU implementations documented in Ref.~\cite{Krause_2025} by orders of magnitude, especially at batch = 1, where the FPGA’s low-latency and deterministic execution provides a clear advantage.

\begin{table}[h]
\centering
\begin{tabular}{ccc}
\toprule
\textbf{FPGA Resource / Latency} & \textbf{Value} & \textbf{Utilization}\\ 
\midrule
Latency [$\mu$s] & $12.29 \pm 4.56$ & --- \\
LUTs              & 1470513           & 85\% \\
FFs               & 437455            & 12\% \\
DSPs              & 1936              & 15\% \\
\bottomrule
\end{tabular}
\caption{FPGA resources utilization (expressed in LUTs, DSPs, and FFs) and latency for the FPGA model. Utilization quantifies how much of the target FPGA's resources are used. }
\label{tab:results}
\end{table}

\begin{figure}[tbh!]
\centering
\includegraphics[width=0.9\textwidth]{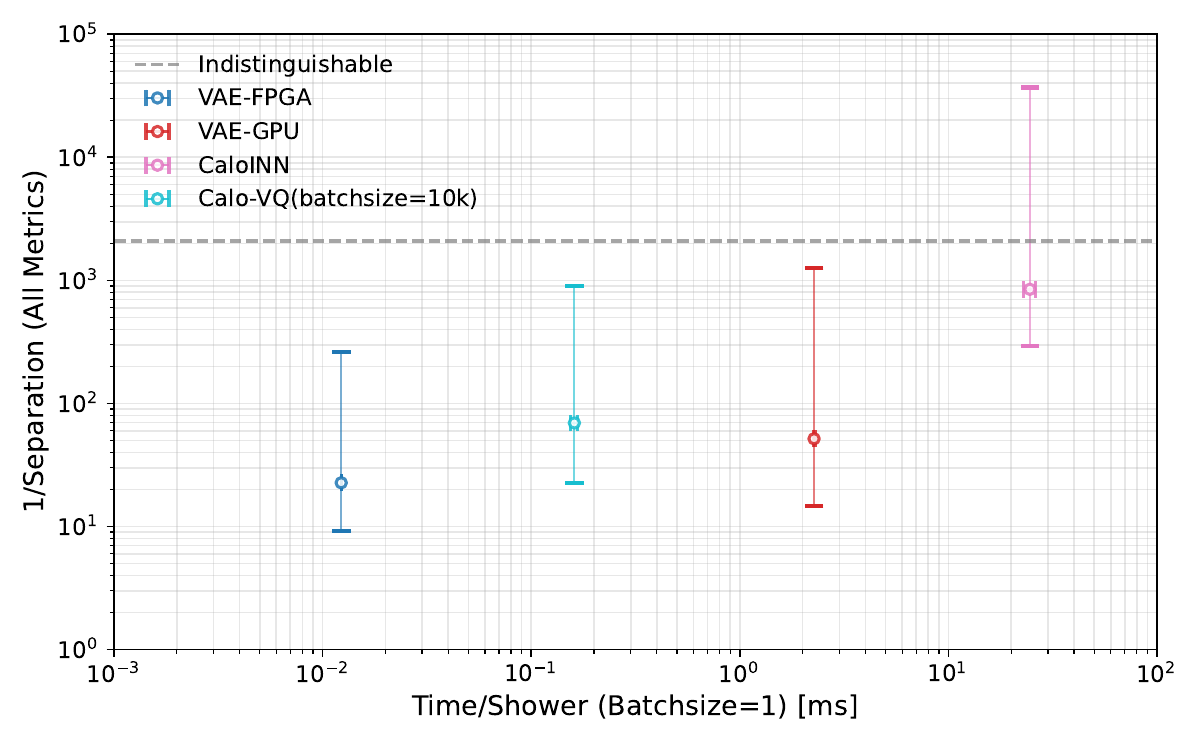}
\caption{Per shower generation speed vs. performance (inverse of average of all separation metrics), comparing the VAE-GPU and VAE-FPGA to two GPU models from the CaloChallenge representing state-of-the-art performance (\texttt{CaloINN} for fastest GPU batch size 1 and \texttt{CaloVQ} fastest at batch size 10$^4$.). 
The error bars correspond to the minimum and maximum separation metrics obtained across all physics observables listed in Table~\ref{tab:v}.
\label{fig:perfVsSpeed} }
\end{figure}

Figure~\ref{fig:perfVsSpeed} provides a summary of these results by comparing the performance-latency trade-off for the VAE model running on either a GPU or FPGA platform, with exact values also listed in Table~\ref{tab:results_speed}.  
Two reference models from CaloChallenge, namely the \texttt{CaloINN} (at batch size of 1) and \texttt{CaloVQ} (at batch size of 10,000) are included to represent the fastest and peak achievable performance, but with the restriction that their size and complexity only affords GPU operation. 
As currently shown, there remains a performance gap relative to larger GPU models in Ref.~\cite{Krause_2025} which have over an order of magnitude more parameters than the model discussed here. 
However, the reduction in latency can provide a meaningful compensation to the simulation quality drop in the context of certain applications, where quantity of simulated showers may be more important than exact precision. 
Furthermore, FPGA deployed algorithms will require less overall power, providing a meaningful reduction of cost and environmental impact, both of which are key priorities for future scientific programs.

To realize these results in practical simulation contexts, future work must focus on further model compression and FPGA implementation strategies to preserve performance while maintaining the efficiency benefit. More compact and higher fidelity model architectures will also be explored to improve simulation quality to experimental standards whilst remaining within FPGA resource constraints. Notably, the presented model operates near the LUT capacity of the target FPGA, indicating the need for more efficient architectures and implementation techniques to enable higher-fidelity models.
Altogether, these results demonstrate strong feasibility of FPGA-based assistance in processing large generative ML simulation loads for HEP applications.


\begin{table}[h]
\centering
\begin{tabular}{lccc}
\toprule
\textbf{Model (Device)} & \textbf{Trainable Parameters} & \textbf{[ms]/shower} & \textbf{$1/S_\mathrm{avg}$} \\
\midrule
VAE-FPGA (FPGA) & 234{,}884       & $0.0122 \pm 0.005$  & 22.7  \\
VAE-GPU  (GPU)  & 234{,}884      & $2.2700 \pm 0.0234$ & 51.6  \\
\texttt{CaloINN}  (GPU)  & 18{,}821{,}350 & $24.6 \pm 1.6$      & 846.5 \\
\bottomrule
\end{tabular}
\caption{Model comparison of per-shower generation speed and performance, quantified by the inverse of the average separation metric ($1/S_\mathrm{avg}$), at batch size 1. Time and $S_{\text{avg}}$ values correspond to those shown in Figure~\ref{fig:perfVsSpeed}.}
\label{tab:results_speed}
\end{table}



%% file: sections/conclusions.tex
\section{Conclusions}
\label{sec:conclusions}

This study demonstrates that a simple generative model, namely a hardware-aware compressed variational autoencoder, can perform fast calorimeter simulation on a single FPGA. 
It achieves a two orders of magnitude reduction in latency for small batch sizes with modest resource usage, while maintaining only minor performance degradation ($\mathcal{O}$(10)\%) relative to GPU-based implementations.
This work provides the first demonstration that existing FPGA resources at LHC experiments can be used for simulation generation during non-data-taking periods. 
These findings indicate a practical path for more power-efficient simulation workflows scalable to future experimental needs, while supplementing conventional workflows by making use of available on-site computing capacity. This study also explores a practical workflow for deploying general ML models on low-latency devices such as FPGAs, enabling potential applications to a wider range of HEP offline tasks, including reconstruction and data processing, through heterogeneous computing architectures with FPGA acceleration and straightforward integration into existing software systems. 

\clearpage

%% file: sections/app_hls4ml.tex
\section{\hlsforml~ Details}
\label{app:hls4ml}

\begin{table}[t]
\centering
\small
\begin{tabular}{p{0.30\linewidth} p{0.66\linewidth}}
\toprule
\textbf{Layer / component type} & \textbf{Precision settings (fixed-point)} \\
\midrule

Dense -- hidden layers &
\begin{tabular}[t]{@{}l@{}}
input/result: ap\_fixed$<$16,6$>$ \\
weight: ap\_fixed$<$6,2$>$ \\
bias: ap\_fixed$<$8,3$>$ \\
mult: ap\_fixed$<$18,8$>$ \\
accum: ap\_fixed$<$20,8$>$
\end{tabular} \\
\addlinespace[0.8em]

Dense -- feeding into layer energy ratio activation &
\begin{tabular}[t]{@{}l@{}}
input/result: ap\_fixed$<$16,6$>$ \\
weight: ap\_fixed$<$8,3$>$ \\
bias: ap\_fixed$<$10,3$>$ \\
mult: ap\_fixed$<$20,8$>$ \\
accum: ap\_fixed$<$28,12$>$
\end{tabular} \\
\addlinespace[0.8em]

Dense -- feeding into energy response ratio activation &
\begin{tabular}[t]{@{}l@{}}
input/result: ap\_fixed$<$16,6$>$ \\
weight: ap\_fixed$<$16,6$>$ \\
bias: ap\_fixed$<$16,6$>$ \\
accum: ap\_fixed$<$42,22$>$ \\
result: ap\_fixed$<$42,22$>$
\end{tabular} \\
\addlinespace[0.8em]

BatchNorm &
\begin{tabular}[t]{@{}l@{}}
input/result: ap\_fixed$<$16,6$>$ \\
mean: ap\_fixed$<$10,5$>$ \\
var: ap\_fixed$<$12,7$>$ \\
gamma: ap\_fixed$<$8,3$>$ \\
beta: ap\_fixed$<$8,4$>$ \\
mult\_t: ap\_fixed$<$18,8$>$ \\
scale\_t/bias\_t/mean\_t/var\_t: ap\_fixed$<$20,8$>$
\end{tabular} \\
\addlinespace[0.8em]

LeakyReLU &
\begin{tabular}[t]{@{}l@{}}
input/result: ap\_fixed$<$16,6$>$ \\
slope\_t: ap\_fixed$<$12,6$>$
\end{tabular} \\
\addlinespace[0.8em]

Softmax &
\begin{tabular}[t]{@{}l@{}}
input/data/result: ap\_fixed$<$16,6$>$ \\
table/exp\_table\_t/inv\_table\_t: ap\_fixed$<$18,8$>$ \\
sum/accum: ap\_fixed$<$20,8$>$
\end{tabular} \\
\addlinespace[0.8em]

Sigmoid &
\begin{tabular}[t]{@{}l@{}}
input: ap\_fixed$<$42,22$>$ \\
result: ap\_fixed$<$16,6$>$ \\
table type: ap\_fixed$<$18,8$>$
\end{tabular} \\
\addlinespace[0.8em]

Concatenate &
\begin{tabular}[t]{@{}l@{}}
input/result: ap\_fixed$<$16,6$>$
\end{tabular} \\

\bottomrule
\end{tabular}
\caption{Layer-type precision settings used in hls4ml. All fixed-point types use rounding mode \texttt{AP\_RND\_CONV} and saturation mode \texttt{AP\_SAT}. Reuse factors are layer-specific for dense layers and are omitted here for brevity.}
\label{tab:hls4ml_layer_precisions}
\end{table}